# Slotted ALOHA on LoRaWAN - Design, Analysis, and Deployment

**Tommaso Polonelli [1], Davide Brunelli [2], Achille Marzocchi [1] and Luca Benini [1,3]**

[1]  DEI, University of Bologna; {tommaso.polonelli2, luca.benini}@unibo.it
[2]  DII, University of Trento; davide.brunelli@unitn.it
[3]  IIS, ETH Zurich; luca.benini@iis.ee.ethz.ch

**Abstract:** LoRaWAN is one of the most promising standards for long-range sensing applications. However, the high number of end devices expected in at-scale deployment, combined with the absence of an effective synchronization scheme, challenge the scalability of this standard. In this paper, we present an approach to increase network throughput through a Slotted-ALOHA overlay on LoRaWAN networks.  To increase the single channel capacity, we propose to regulate the communication of LoRaWAN networks using a Slotted-ALOHA variant on the top of the Pure-ALOHA approach used by the standard; thus, no modification in pre-existing libraries is necessary. Our method is based on an innovative synchronization service that is suitable for low-cost wireless sensor nodes. We modelled the LoRaWAN channel with extensive measurement on hardware platforms, and we quantified the impact of tuning parameters on physical and medium access control layers, as well as the packet collision rate. Results show that Slotted-ALOHA supported by our synchronization service significantly improves the performance of traditional LoRaWAN networks regarding packet loss rate and network throughput.

**Keywords:** Internet of Things; Wireless Sensor Networks; Long Range Radio, LoRa, LoRaWAN, ALOHA, Slotted ALOHA, Synchronization, Low-energy wireless protocols.

## 1.  Introduction

In past years, industry and academia have dedicated significant efforts to develop low-power wide area networks (LPWAN), as a new category of wireless communication standards [1][2]. Among commercial protocols, LoRaWAN, and the LoRa modulation developed by Semtech, is gaining popularity [3-11]. The public availability of the specifications [12] combined with low-cost certified transceivers [13] and the fact that a LoRaWAN WSN can operate in the unlicensed radio spectrum, means that anyone can use the radio frequencies without having to pay million dollar fees for transmission rights, motivates the global attention on this protocol. LoRa offers a link range up to 10km with a single gateway, which is capable to support hundreds of thousands of different devices; moreover, a unique feature of LoRa is the ability of the gateway to demodulate simultaneously multiples transmissions, with different data rates, on the same channel and time. LoRaWAN specification describes three classes of operations: low-power bi-directional end devices (Class A), scheduled downlink transmission (Class B) and Class C designed for always-on bi-directional actuators. In this work, we focus the attention on the most common used Class A.

Despite the initial success, multiples issues remain open about the real performance in terms of scalability and the maximum traffic supported in massive LoRaWAN installations. Most LPWAN protocols promise to manage thousands of devices, but the impact of the LoRaWAN parameters on large-scale networks, with packet time that can be more than one second, is still not well explored. Indeed, in [14], [15] and [16] the scalability of LoRaWAN is studied, and the results in [14] show that only 120 nodes can be managed with a packet transmission every 22 minutes with static configurations. The performance improves growing up to 1600 with dynamic configurations that minimize the time of air. In [5], the authors show an impressive reduction of channel capacity, up to 2x, when the acknowledgments of the messages are required. However, these figures are not enough for future IoT deployments [5]. While many existing works have studied the scalability of LoRaWAN networks, most of them do not consider the impact of downstream traffic in channel throughput, interference, and issues related to real deployments, such as clock drifts and delays produced by restricted computational resources of LoRaWAN gateway and server.

The work presented in this article aims at providing an analysis of the channel throughput of the LoRaWAN protocol, and to investigate the issues associated with the scalability. The results can provide useful guidelines for large-scale deployments and physical configurations. A specific network simulator offers information about the real performance of LoRaWAN Class A with varying number of devices and packet time of air. We studied the effective payload throughput related to the protocol overhead, which decreases with a factor of 2.22x (in the worst case) compared to Pure-ALOHA (P-ALOHA). After that, we propose a solution to improve network performance based on Slotted-ALOHA (S-ALOHA). We developed a reliable synchronization algorithm optimized for low-power devices, with an accuracy of 5.37ms and a success rate over 99%. This reference time is used to keep aligned all the end nodes in LoRaWAN network, allowing deployment of a Slotted LoRaWAN built on top of standard LoRaWAN specifications. Simulations and tests, performed with 24 end nodes, show a throughput improvement of 2x, and 5.8x in crowded conditions. The low-power capabilities of Class A are preserved, and our Slotted LoRaWAN can be implemented on existing commercial products.

The remainder of this work is organized as follows. Section 2 presents the related works, while section 3 introduces the LoRa physical layer and the LoRaWAN protocol. Section 4 models the channel throughout in the worst case and verifies the statistical analysis with results generated from a network simulator. Section 5 provides the description and of the synchronization algorithm and section 6 shows the implementation of the S-ALOHA in LoRaWAN network. Finally, section 7 and 8 investigate the Slotted LoRaWAN throughput and synchronization errors. Section 9 concludes the paper.

## 2. Related Works

The number of connected IoT devices is expected to rise at an annual rate of 32% and market forecasts expect 21B devices by the end of this decade [2]. Some of the newest applications require to cover entire cities or huge buildings at very low energy use and cost. These requirements are hard to reach using traditional methods and infrastructures such as WPAN or cellular [1]. Large-scale IoT installations are becoming a reality, as networks are being deployed for urban monitoring applications [2], smart city [18], intelligent transportation systems [19], urban monitoring applications [20]. Several radio technologies, such as Sigfox, LoRa™, IEEE 802.15.4 NB-IoT, BLE 5.0 and DASH7, are currently competing in the arena of Device-to-Device (D2D) long-distance, low-power communication [21].

Different communication technologies aimed at low power have been proposed and deployed. It is possible, as discussed in [5], to split them in two categories: low power local area networks, such as IEEE 802.15.4, Bluetooth Low Energy which are typically utilized in short-range personal area networks; low power wide area networks, with a transmission range greater than 1000m; this category includes LoRaWAN, Sigfox, and DASH7 as major players. Sigfox is a protocol based on a variation of the cellular system which permits remote devices to connect to an access point through Ultra Narrow Band. It is a proprietary technology, developed and distributed by Sigfox. Full specifications are difficult to obtain, so it is difficult to carry out a clear and comprehensive comparison with other protocols. Each end-device can send 140 messages per day, with a payload size of 12 octets, at a data rate up to 100 bps. Each Sigfox access point can handle up to a million end-devices, with a coverage area of 30–50 km in rural areas and 3–10 km in urban areas [22]. The limit on payload and number of packets makes Sigfox networks hardly usable in application scenarios where communication is not very sporadic: LoRaWAN is less prescriptive and more flexible.

DASH7 [23] is a wireless sensor and actuator full Open Systems Interconnection (OSI) stack protocol that operates in the 433MHz, 868MHz, and 915MHz unlicensed ISM band. It originates from the ISO 18000-7 standard [24]. DASH7 can provide communication in the range of 2km with low latency, mobility support, and multi-year battery life. A secure AES 128-bit shared key encryption is supported with a data rate up to 167Kbit/s. However, DASH7 needs more power per bit than other protocols like LoRaWAN, and for applications that do not require synchronization or low latency uplink, LoRaWAN would be preferred. Moreover, during fast motion in rural areas, DASH7 connections are not as reliable as other LPWAN solutions.

In LoRaWAN™ architecture [25], sensor nodes communicate with the gateway, which serves as a bridge between the nodes and a network server. Three types of functional classes are defined for end-devices: A, B, and C. Currently, several papers analyze LoRa performance [5-30]. In these works, LoRa deployments are compared in terms of network throughput and power consumption. In [3] the coverage of LoRa is studied and in [31], the authors analyze the LoRa network capacity and propose LoRa-Blink to support multi-hop communications. In [32], Casalas and Mir present an analytical model that characterizes the current consumption, lifetime and energy cost in LoRaWAN Class A. The proposed model allows to quantify the impact of physical and MAC layers, as well as bit error rate and collisions, on energy performance.

As in the [25] Class A ISO/IEC ISM regulations, the end nodes and gateways can transmit "at will" without any carries sensing. This makes the LoRaWAN's MAC similar to ALOHA [33]. The ALOHA mechanism can cause substantial inefficiency in LoRaWAN networks, as highlighted by the simulation study reported in [26], which confirmed that increasing the number of gateways can improve the global performance but cannot eliminate fast saturation. Indeed, generating packets with Poisson law and uniform distribution of the payload lengths between one and 51 bytes, as simulated in [5], the maximum expected channel capacity usage is 18%, which is similar to the Pure-ALOHA (P-ALOHA) throughput, with a link load of 0.48. This may be an issue because several packets could be lost; indeed, at the maximum channel load, around 60% of the packets transmitted are dropped because of collisions. To verify the correct reception of each packet an acknowledgment (ACK) method could be applied, but it requires two successive transmissions in order to be successful, the payload in uplink and ACK in downlink, increasing the collision probability with other messages.

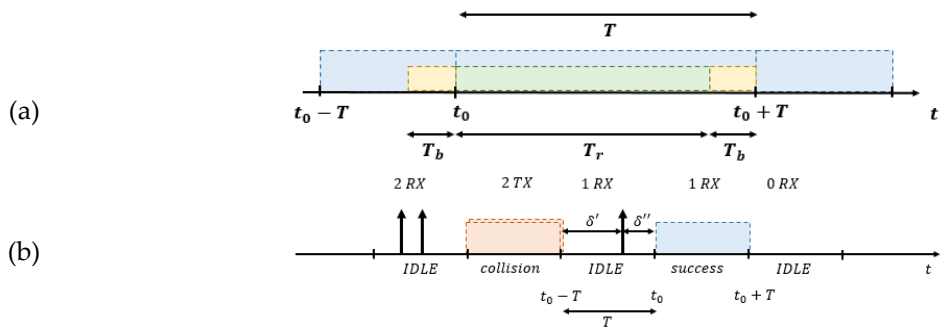

**Figure 1** (a): Slot width definition in S-ALOHA, (b): Example of free, collided and successful slot

S-ALOHA protocol is well known in local wireless communications [34] since the 70's. In S-ALOHA the channel time is divided into slots (Fig. 1), which have fixed length T and are basically composed of two parts: transmission time ($T_t$) and the tolerance interval ($T_b$), as presented in Figure 1 (a). Every end node must transmit a packet only at the beginning of a slot. If two or more end nodes transmit their packet during the same period, a collision occurs; otherwise, no collision is generated, and the data are properly sent (Figure 1 (b)). If no terminals access in one slot, it will be free, and no device will start the transmission in the middle of the time-slot. The maximum theoretical S-ALOHA channel throughput is 37%.

## 3. LoRa physical layer and LoRaWAN

LoRa is a chirp spread spectrum modulation. It is a signal which frequency increases or decreases with time. It was developed for radar applications; indeed, these signals have a constant amplitude and pass the whole bandwidth in a linear or non-linear way from one end to another end in a defined time. The frequency delta allowed between the receiver (gateway) and transmitters (end nodes) can reach 20% of the bandwidth without impacting decoding performance [5]. This factor is useful to design cheap devices with low-quality crystals, such as commercial products with 80 ppm/°C. LoRa typical transceiver, such as the SX1276 [35], offers a link budget up to 168 dB with a receiver sensitivity of -148dBm. Since the symbol period is longer than a typical noise spike, errors generated from these interferences are easily filtered at the receiver side, through error correction codes. Code Rate (CR), Spreading Factor (SF) and Bandwidth (BW) parameters affect the LoRa modulation; indeed, a LoRa symbol is composed of two SF chirps, with a frequency variation that covers the entire band. In LoRa modulation the chirp symbol rate is directly dependent on the bandwidth, this parameter generates different consequences as specified in [5]: (i) increasing the SF, the chirp frequency span is divided by two; (ii) the period of each symbol is multiplied by two; (iii) since more bits are transmitted in a symbol, the bit rate is not decremented with a two ratio; (iv) the bit rate at given SF is proportional to the bandwidth, doubling the bandwidth will double the transmission rate. Over LoRa modulation, a forward error correction is used to enhance the communication reliability. This method is implemented through four different coding rate (CR) approaches, denoted CR1 to CR4. Lastly, a low data rate drift correction mechanism is applied to increase robustness to frequency variation over the timescale of the LoRa message. Equation 1 gives an estimation of the equivalent bit rate (EBR), ), detailed information can be found in Supplementary Materials and Appendix A:

(1) $$EBR = SF\left(\frac{BW}{2^{SF}}\right)CR$$

The radio channels used in a LoRa deployment depend on the country rules. In this paper, all the tests are performed in Italy, Europe, where the ISM (Industrial Scientific Medical) band is at 863-870MHz. LoRaWAN provides 9 different channels, channels from 0 to 9 have a bandwidth of 125KHz and must support data rates between 0.3kbps to 5kbps; indeed, channel ten is allocated for FSK modulation with a bandwidth of 250KHz. Figure 2 presents the frequency characterization of each LoRaWAN channel.

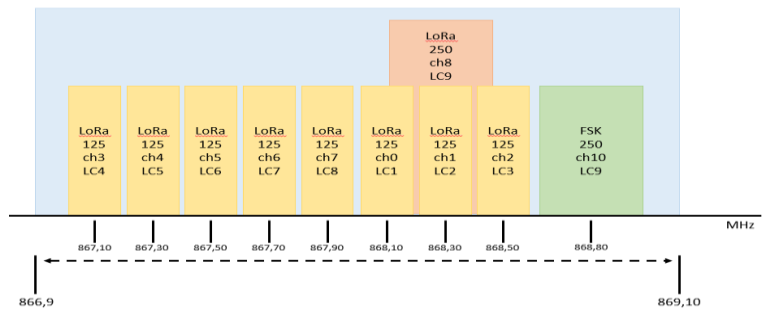

**Figure 2 -** Description of LoRaWAN channels

An important feature of LoRa modulation is that the different SFs are orthogonal, this means that a LoRaWAN gateway can receive a multiple transmission on different SFs simultaneously and can apply this methodology for each channel.

Due to local rules imposed by ISO/IEC ISM regulations, wireless sensor devices working on ALOHA MAC access cannot occupy more than 1% of the channel time. It is important to underline that, as long as the limitations for each band are respected, each end device can transmit on different channels contained in different sub-bands in order to increase its overall throughput [9].

Class A in LoRaWAN is focused on the end node side; indeed, all transmissions are scheduled by the end node, whereas the server can transmit only in one of two receive windows opened after a previous uplink transmission, as presented in Figure 3. Therefore, any packet that the application level needs to transmit must wait until the next scheduled receive windows.

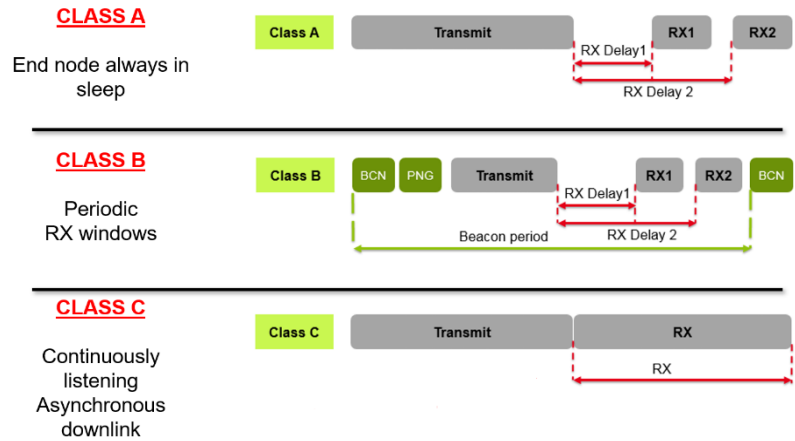

**Figure 3 -** Uplink and Downlink methodology in Class A, B, and C

In the MAC layer the typical data packet is composed by a preamble, a physical header, a variable payload length and, lastly, error detection bytes (CRC); additionally, the CRC is present only in the uplink packet, thus the downlink data is often used for ACKs. LoRaWAN MAC defines different strategies for the downlink channel and DR selection in RX1 and RX2 windows; in this paper, we set up the WSN to use the same DR and channel for uplink and downlink transmissions; moreover, we disable the RX2 window to evaluate the network capabilities with static conditions. Indeed, with the RX2 enabled, the overall downlink period could change up to 200%. The minimum time width for both receiving windows must guarantee at least the time required to effectively detect a downlink preamble [25], and if preamble symbols are detected, the transceiver must remain active until the reception of all data.

## 4. LoRaWAN MAC Model

Class A is designed for battery-powered end nodes that must be ultra-low power with very long expected lifetime. It permits bi-directional communications, but downlink transmissions are constrained to two short intervals after each uplink transmission to the gateway. The MAC of LoRaWAN is based on P-ALOHA. End-devices of Class B and Class C are designed for end nodes with no critical energy constraints, thus they allow for more receiving slots at specific scheduled times and synchronized beacons are sent from the gateway. In this paper, we focus on an S-ALOHA scheme for LoRaWAN battery-powered sensor nodes.

We define S as the average number of packets generated per transmission time interval; the traffic source $\lambda$ consists of a high number of users who collectively form an independent Poisson source with an aggregate mean packet generation rate of X packets/s, the packet time width is supposedly fixed with a period of T seconds (Equation 2). Moreover, we can consider that each user generates packets infrequently and that each packet can be successfully transmitted in an interval smaller than the average time between successive packets from a given node [36]. S can also be expressed as the channel throughput rate. Each node delays the transmission of a previously collided packet by some random time, chosen, for example, uniformly between 0 and $T_{max}$. Therefore, the traffic injected into the channel consists not only of new packets but also of previously collided packets: this increases the mean traffic generated, usually denoted with G (Equation 3).

$$S = \lambda T \tag{2}$$

$$G \geq S \tag{3}$$

$$G(n) = \lambda(n)T \tag{4}$$

$$S = G(n) \cdot P_{suc} = \lambda(n)T \cdot e^{-\lambda(n)2T} \tag{5}$$

In P-ALOHA, a single transmission is successfully performed if during the time period 2T (vulnerability period) the channel remains free. The probability that there are no transmissions in the 2T period is $P_{suc}$ [36]. The total channel traffic could be expressed as presented in Equation 4. With these hypotheses, it is possible to obtain the maximum channel throughput: 18%.

This model generates results that approximate to real deployments for LoRaWAN Class A end node in the unacknowledged configuration when there is only uplink transmission without the answer from gateway(s). To realistically model the single channel throughput in the half-duplex mode we achieved a statistical model based on measurements from a real LoRaWAN testbed, the network simulator and statistical analysis.

As shown in Figure 3, in LoRaWAN Class A there is one second delay between transmission and RX1 windows; this period could be used by other end nodes to transfer data, but the probability of a successful transmission in the RX Delay 1 is very low, since LoRa modulation generates a remarkable packet's time of air. With an SF 12, a BW 125 KHz, CR 4 and 25 bytes of payload the overall time of air is 1253ms, with these setting the corresponding ACK time width is 530ms.

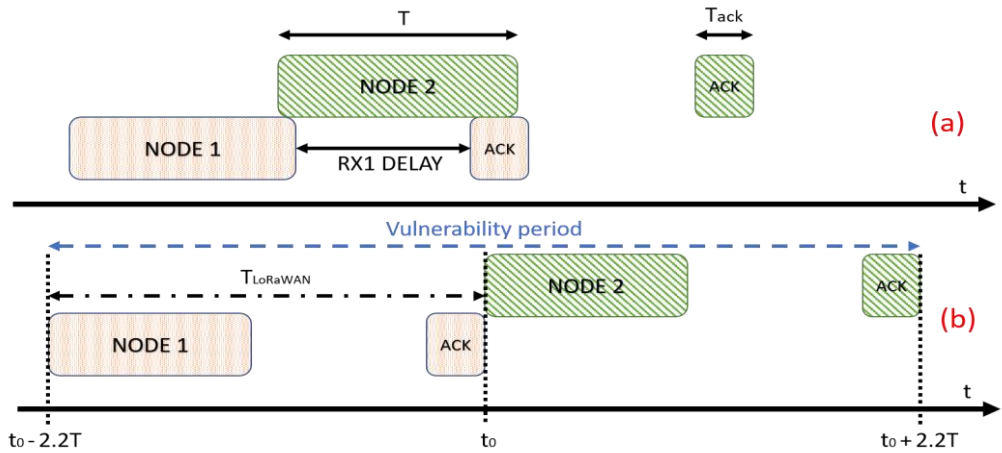

**Figure 4 -** Vulnerability period in LoRaWAN with ACK. (a) T is greater than RX1 Delay, a collision must occur; (b) vulnerability period needed to perform a successful link between an end device and the gateway.

With the proposed parameters, a transmission inside the RX1 Delay is not allowed, as shown in Figure 4, and the probability of collision is equal to one. In the proposed model the overall channel occupation time for each transmission is composed by time of air of both uplink (T) and downlink ($T_{ack}$) windows plus the RX1 Delay. If we consider T the channel time used to transfer the payload, we can define 2.22·T the total channel time used for each sensor, considering the LoRaWAN overhead; this value is equal to $T_{LoRaWAN}$. Therefore, the ratio between the traffic injected and transmission successfully performed can be expressed as presented in Equation 6, where the vulnerability period scales up from 2T to 2·(2.22T).

$$ (6) \qquad S = G(n) \cdot P_{suc} = \lambda(n)T \cdot e^{-\lambda(n)2(2.22T)} $$

The effective LoRaWAN throughput decreases drastically with respect to the classic P-ALOHA, with a maximum of 8% at G equal to 0.25. This result shows that the LoRaWAN throughput (for each channel) is poor when half-duplex communication is required. To confirm this conclusion, we developed a simulator in MATLAB to evaluate the channel collision rate in a WSN with different time-of-air configurations with the uplink and RX1 windows. In details, the simulator generates traffic on a variable number of nodes and each one transmits following a Poisson uniform distribution. When a simulation starts, the number of nodes grows from 1 and stops as soon as the desired (G) is achieved. The maximum number allowed is 1000, and each element tries to access into the channel with a variable duty cycle and a maximum of 40 attempts. The simulator checks every 1ms if two end devices try to communicate simultaneously, with an overall simulation time of two hours. Figure 5 shows the S(G) for unconfirmed and confirmed LoRaWAN single channel link; indeed, in Figure 6 (a) and Figure 6 (b) the results of two different simulations are presented. In Figure 6 (a) the packet time is greater than the RX1 Delay, in Figure 6 (b) the time of air is << 1 second, so the probability for a transmission successfully performed is not equal to zero, and the network performance slightly increases.

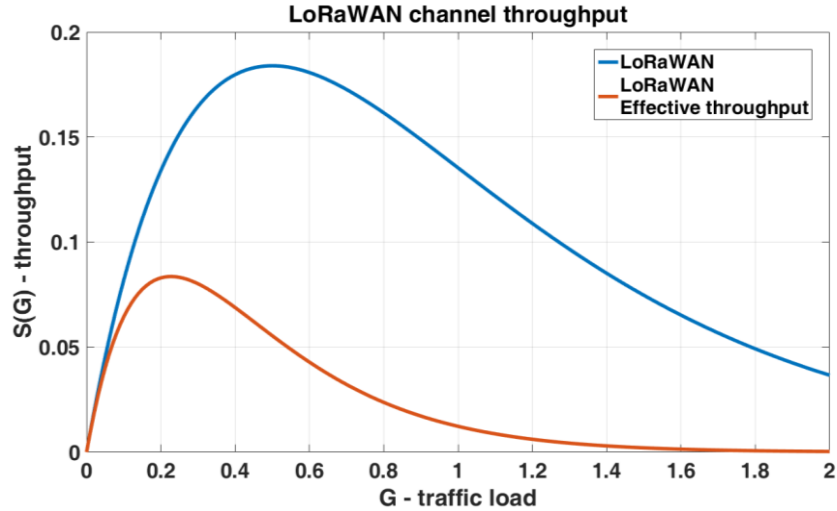

**Figure 5 -** Unconfirmed and confirmed LoRaWAN single channel throughput

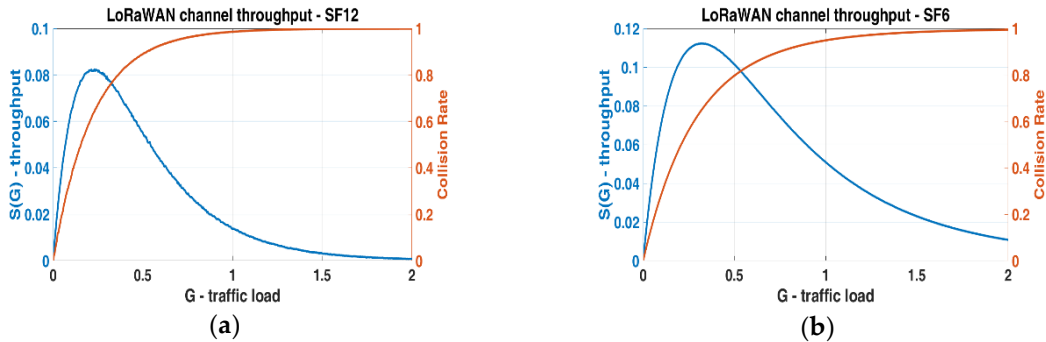

**Figure 6 –** Confirmed LoRaWAN throughput and collision rate for one channel acquired by MATLAB simulation. (a) SF=12, BW=125KHz, CR=4 Maximum throughput is 8%; (b) SF=6, BW=125KHz, CR=4 Maximum throughput is 11%.

Figure 5 and Figure 6 show that the proposed model is valid and that the effective channel throughput in LoRaWAN networks is tightly constrained. In S-ALOHA MAC protocol all the components of a LoRaWAN link can be fitted into a single slot (Figure 1 (a)), uplink and downlink windows as well as the RX1 Delay, keeping the channel throughput of Slotted LoRaWAN (S-LoRaWAN) equal to standard S-ALOHA. As expressed for the P-ALOHA, if we consider T the channel time used to transfer the payload, we can define 2.22·T the total slot time used, taking into account the LoRaWAN overhead. With these considerations, the maximum channel throughput in S-ALOHA is 16%, doubled respect to the classic half-duplex LoRaWAN.

## 5. Real Time Clock Synchronization over LoRaWAN stack

Clock synchronization across end nodes is fundamental to define the slots used by S-ALOHA. In literature, articles experimented some synchronization schemes on the top of the bare LoRa radio [28], without any LoRaWAN protocol to demonstrate that LoRa modulation can be used even industrial environments [29]. Here we use a lightweight synchronization methodology, presented in [27] and tailored for LoRaWAN end nodes and successfully tested on an STM32L4 MCU. To generate a time reference for the S-ALOHA protocol, where the slots must be kept aligned in all the end devices, we use a real-time clock (RTC) with an inexpensive external crystal at 32.768KHz with a thermal variability between 20 to 80ppm/°C.  Since this component drifts significantly with temperature, we need to re-synchronize end node clocks frequently to keep slots sufficiently aligned. The synchronization scheme is piggybacked on the predefined answer (ACK) of a LoRaWAN Class A RX1 window, which is opened with a maximum error of ±20µS delay [27]. Common events for both devices, end node and gateway, are used as a shared reference for the synchronization algorithm. In a typical half-duplex link, the node and the gateway must know when to open the RX1 window after a successfully uplink transmission. A timestamp at the end of such packet is saved on both devices with the goal to open the RX1 window simultaneously; thus, this event (Figure 7) is used as the reference for our clock synchronization procedure [27].

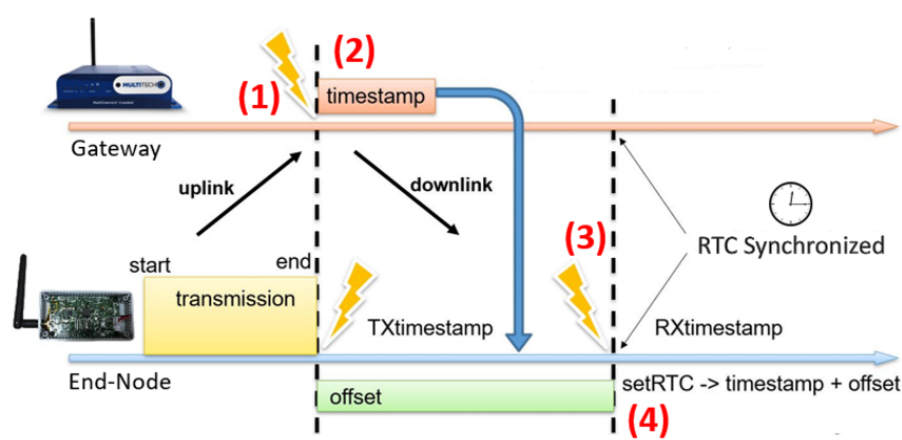

**Figure 7 -** Synchronization procedure

The synchronization algorithm is based on a clock reference distribution system: basically, at the end of every complete uplink transmission each node updates its real-time clock. We implemented an algorithm that works as follows: (1) gateway and node both save a timestamp of the moment when the uplink transmission has ended; (2) the node receives, in the first available RX window, gateway's timestamp piggybacked in the acknowledgment message; (3) due to the non-negligible flight times and the RX window's delay, the node must consider the offset between the timestamp when the uplink transmission has ended (TX timestamp) and the one when it receives the acknowledgement message (RX timestamp). Node calculates the difference between these two timestamps and adds it to the timestamp received by the gateway. (4) The ode updates its real-time clock with the new timestamp, in this way both devices are synchronized.

The developed system must automatically update the end node RTC to guarantee an adequate alignment with the gateway's reference clock. Moreover, the protocol must be compliant with LoRaWAN specification without adding a significant overhead on the overall communications. The proposed approach uses the LoRaWAN uplink packet and the RX window already scheduled after 1 second; indeed, if the timestamp saved on gateway, received at the end of the uplink transmission, can be sent back with the ACK packet in the RX1, the end node has the time reference needed for a clock synchronization. Overhead is very low, only 8 bytes of timestamp, with good scalability and configurability [27]. In many applications the overhead is negligible, for example when the end node generates many data packets, the synchronization comes for free adding the gateway timestamp at some downlink packet. In WSN where the end node is programmed to send few packets every day, some additional synchronization phases will be needed to ensure clocks do not drift too much. Maximum tolerable clock drift depends on three factors: (i) maximum uplink time of air; (ii) daily duty cycle of each end node; (iii) maximum throughput required for the specific application.

On the end node side, the purpose of the RTC synchronization function, implemented in firmware, is to update the time contained in the local RTC with the timestamp received from the gateway, this procedure requires two steps: the calculation of the offset and the management of the sub-seconds. Since the packet in response to the uplink transmission takes a variable delay before reaching the node, the offset calculation is needed to update the received timestamp that could be no longer current. The offset calculation procedure consists in measurement between the response packet and the initial instant when the transmission by the end node ends, this instant coincides also with the arrival of the packet to the gateway, which is the moment where the gateway "marks" the message with the timestamp included in the RX1. The electromagnetic propagation speed is negligible factor respect to the times involved.

A total of 30 synchronization tests were carried out with 2, 3 and 4 nodes: the results are shown in Table 1, where the time difference in ms is expressed taking as reference the end node N0. The purpose of this evaluation is to verify the synchronization performance with many devices, indeed they are programmed to send a timestamp request every 15 seconds, which is comparable with 320 end nodes with a request-rate of 80 minutes each.

**Table 1 -** Results of the measurements carried out to verify the RTC synchronization between multiples end node connected to a single Gateway

| Number of end nodes | Min. Synch. Error [ms] | Max. Synch. Error [ms] | Avg. Synch. Error [ms] | Delta Avg. Error [%] |
|---|---|---|---|---|
| 2 | 0.072 | 8.296 | 3.310 | - |
| 3 | 0.120 | 33.260 | 4.705 | +42 |
| 4 | 0.243 | 37.080 | 5.370 | +62 |

From Table 1 it is possible to notice how the increase of devices that request the synchronization from the gateway significantly reduces the synchronization accuracy; this result is probably due to the management of processes at the operating system level within the gateway. Since the management of the timestamps is done by the software working on the Linux operating system, the priority of the processes on the gateway is not well known, and consequently, the final accuracy of the timestamp associated with the received packet cannot be under one millisecond. The results achieved shows that the average error between all the devices is 5.37ms, with a maximum value of 37.08ms, such uncertainty must be considered in S-ALOHA deployment to avoid transmission overlaps between adjacent slots. To decrease the timestamp error some LoRaWAN gateways embed a GPS module, used by the physical layer to associate exactly the time reference at the end of the packet frame, but, since this feature is not mandatory from the protocol rules, we tested our Slotted LoRaWAN without this support. Figure 8 (a) presents the plots of synchronization deviation between end nodes, from two to four, where are indicated the median, and the bottom and top edges of the box, respectively the 25th and 75th percentiles. Figure 8 (a) shows that, as the number of nodes increases, the mean and the width of the interquartile gap (blue box) increases, which is a dispersion index provides a measure of how far the points move away from a central value, such as standard deviation. As a result of the increase in the number of devices, the number of "off-average" synchronizations indicated in the graph with red crosses also increases.

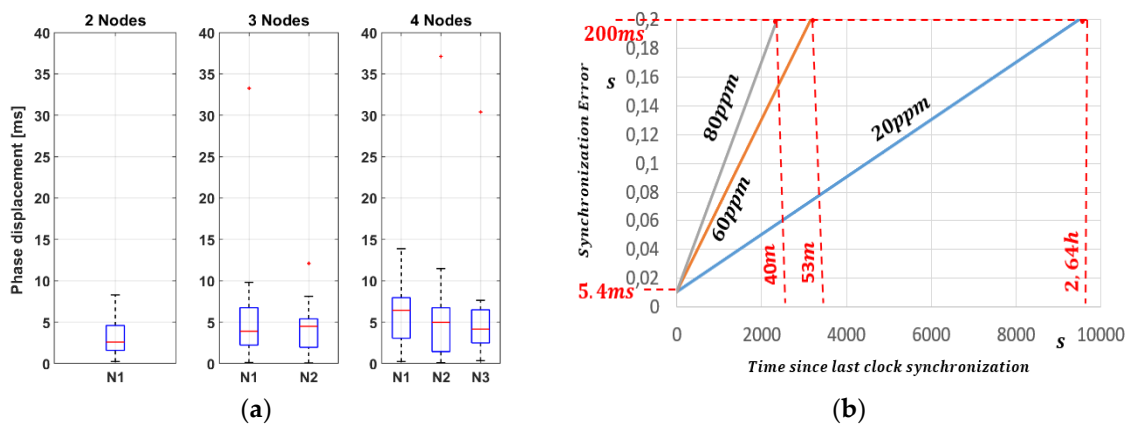

(a)                    (b)

**Figure 8** - (a) median, 25th, and 75th of synchronization error between 2-4 end nodes; (b) Typical crystal drift over time

Figure 8 (b) represents the maximum synchronization uncertainty between nodes since last clock alignment and provides fundamental information to calculate the $T_b$ period as well as the RTC refresh rate, which are dependent from application and transmission period. For example, in the worst case, when the end node integrates an 80ppm crystal, an error of 200ms is generated every 40 minutes.

## 6.  S-ALOHA implementation over LoRaWAN

Once implemented, at the application level, an efficient and accurate synchronization system that guarantees acceptable uncertainties of the order of milliseconds, it is necessary to introduce specific functions for the transmission management on top of LoRaWAN MAC level. On the end node side, the application layer must schedule the uplink transmission only at the beginning of each slot, starting an immediate transfer request when the internal RTC triggers the event. For the calculation of the time $T_r$ it is necessary to consider the fact that the LoRa radios allow different types of configurations depending on the spreading factor, bandwidth, coding rate and, above all, depending on the length of the payload of the packet sent. The configuration used for the tests carried out in the following chapter is made up with these parameters: SF=8, BW=125KHz, CR=1, Preamble=8bytes, and Payload=200 bytes. Following [32], we can estimate the uplink time of air, equal to 553.47ms. This means that considering the 8 bytes added into ACK needed for synchronization procedure, the lower bound of the slot duration is 1.615 seconds ($T_r$), which is equal to $T_{LoRaWAN}$ portrayed in Figure 4. The size of the packets is considerable and close to the maximum allowed of 255 bytes, this to maintain an acceptable ratio between the time used to transfer the data ($T_r$) and the overhead time ($T_b$). This result has been calculated considering that the response in the first window RX1 must occur with the same transmission parameters of the end node. Moreover, it is avoided that the transmission of the ACK message collides with other transmissions being completely included within the time of a slot. The calculation of the $T_b$ time, the tolerance interval, takes place considering the average alignment error introduced by the synchronization algorithm and the clock drift introduced by the quartz oscillator embedded in the microcontroller. Considering a crystal with an error of 80ppm/°C (worst case), a phase shift of 200ms is generated every 40 minutes, a considerable and not negligible value in the case of devices that transmit a few times in a day.   For the calculation of $T_b$, we considered a base-time of 5.4ms given by the maximum alignment shift reached during the synchronization tests to which was added, to consider the clock drift, a time margin of about 400ms in order to have an interval of at least 80 minutes between one synchronization and another. Lastly, a $T_b$ of about 385ms has been chosen, then a slot time T with a total duration of 2 seconds, with a $T_b$ / $T_r$ ratio = 25%.

In addition to the definition of the slot time, to implement the S-ALOHA protocol and make evaluations about the performances and improvements introduced, it was decided to force the transmission of the nodes on a single channel (channel 6). This is to avoid the random selection algorithm of the transmission channel by the LoRa MAC, which would involve the evaluation of the performance on several channels, which goes beyond the scope of this paper.

The S-ALOHA protocol must provide confirmation of the receipt of packets. In case of loss of packets, the end node needs to schedule retransmission in a random slot (random backoff). The backoff algorithm has been implemented in the LoRaWAN application layer, which allows calculating the wait time before retransmitting as a multiple of the single slot time. In the function, a random natural number is calculated within the interval that can be set by the user [0, NSLOTS]. The randomness in the choice of the slot in which to perform the retransmission is a fundamental feature to avoid devices entering pathological repeated interference corner cases.

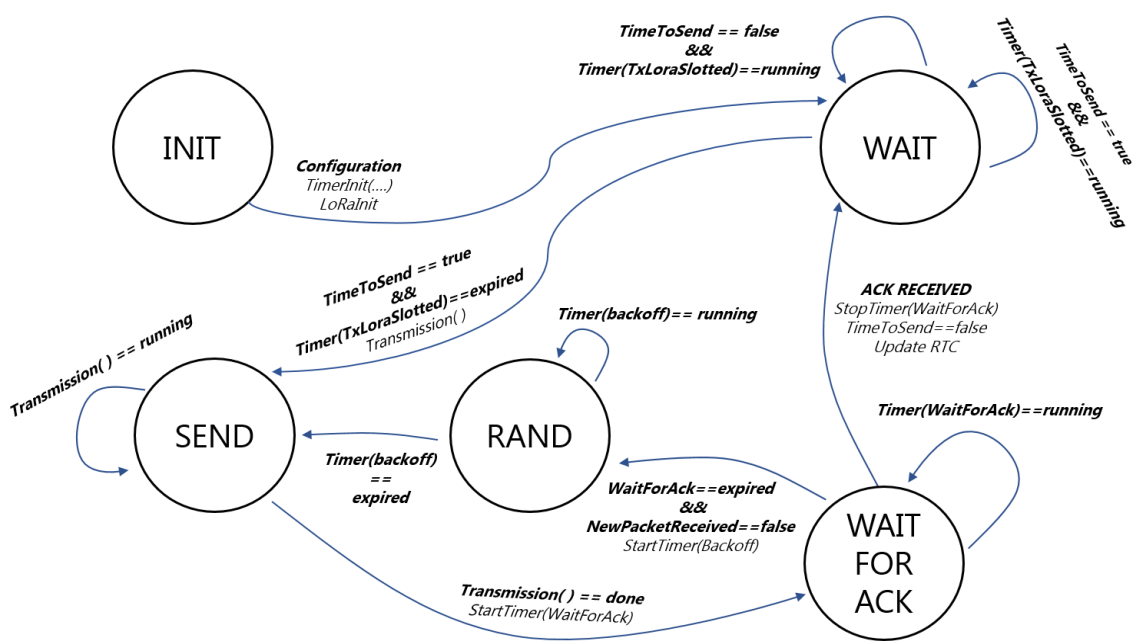

**Figure 9** - S-LoRaWAN block diagram

S-LoRaWAN is implemented at the application layer through the block diagram presented in Figure 9, using two timers and the LoRa transmission libraries. The S-ALOHA state machine has five states: (i) Init: in this state, the timers used during firmware execution are initialized. There are three timers: *TxLoraSlotted* for the calculation of the time to wait to transmit in the first available slot; *TimerBackOff* for the choice of a random slot for retransmission, and *TimerAck* at the expiration of which packet acknowledgement receipt is checked. Following the initializations, the system transitions into the Wait state, in which a transmission request is expected. In this state, the MCU is in sleep. (ii) Wait: *TimeToSend* variable is periodically checked, when it assumes logical value true it is necessary to transmit a new packet to the LoRaWAN network. The state transition between Wait and Send occurs with the triggering of the *TxLoraSlotted* timer in correspondence of the first time slot in which to transmit (as foreseen by the Slotted Aloha protocol). (iii) Send: in this state, the transmission function is executed, which is responsible for sending the data and calling the *PrepareTxFrame* and *SendFrame* functions. These functions deal respectively with the preparation of the payload and the communication with the LoRa transceiver. Following transmission, the system automatically transits to the WaitForAck state. (iv) WaitForAck: the system remains in this state until the timer *TimerAck* is triggered or the acknowledgement is received. If the ACK is received, the system returns to the Wait state until the next transmission request, otherwise the system transitions to the Rand state where the backoff algorithm is executed. (v) Rand: as a result of non-acknowledgment reception, the MCU programs the transmission in a randomly selected slot. The *TimerBackOff* is programmed and the device remains in the Rand state until it expires. When it expires, the system automatically transits to the Send state where the message is retransmitted.

## 7. S-ALOHA evaluation in a real deployment

### a. The end device

The end node used in this paper is based on a custom board developed for multiple purposes. It embeds a STMicroelectronics STM32L476, the LoRa RFM95W transceiver, an energy harvester sub-circuit (BQ25570) that manages the power supply, a temperature and humidity sensor (SHT21). Lastly, an expansion connector enables the board to be connected with several analogue and digital external sensors. The current in sleep mode is 4µA at 3V with the RTC enabled; the STM32L476 uses 8.25mA in RUN mode @48MHz. The RFM95W power consumption in TX and RX is respectively 76 (@10dBm) and 11.5mA. The LoRaWAN firmware comes from I-CUBE-LRWAN libraries package from STMicroelectronics, which is configured to work as our S-LoRaWAN. If the MCU is programmed to send a packet every 30 seconds, the battery (1000mAh) lifespan is seven months. Figure 10 presents a picture of the board and the related block diagram.

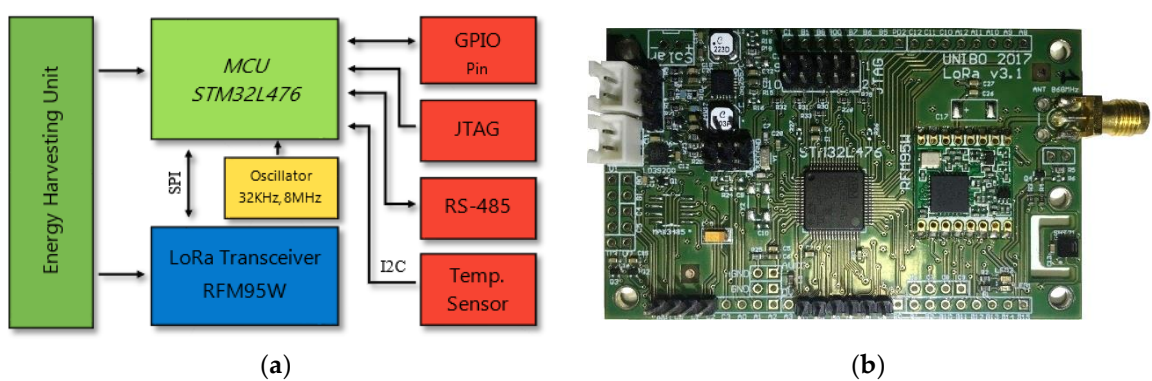

(a)                                                                 (b)

**Figure 10 -** End device HW. (a) Hardware block diagram; (b) Picture of the custom board.

### b. The gateway

The LoRaWAN gateway is a MultiConnect Conduit device. It is a highly configurable gateway for industrial IoT applications. LTE, 3G and 2G, plus Ethernet, are available to deploy network connections and data management. The LoRaWAN radio module embeds the Semtech SX1301 and two SX1257, which demodulate the packets received simultaneously on all channel and SFs. In our deployment, the gateway is connected through ethernet in LAN with the server, and the application software runs on top of Node Red [38].

### c. The environments

We tested our custom sensor node in different deployments and applications. The results presented in this paper are acquired placing the network indoor, allowing an easy setup and configuration management for all the tests presented in the following chapters. We tested our S-LoRaWAN in CINECA (Figure 11), the Tier0 supercomputing center for scientific research in Italy, where our end-nodes were used to monitor in a distributed fashion the temperature evolution of the Computer Air Conditioners with the goal to improve the cooling efficiency. We demonstrate [27] a packet collision reduction of 3.4x in comparison with a standard LoRaWAN WSN with 20 end-nodes. Moreover, we tested our S-ALOHA in a structural monitoring application with a custom sensor node developed to measure the width of cracks in building walls [17]. This application would require hundreds of sensors, which generate thousands of bytes per day. This traffic can overcrowd a standard LoRaWAN network; indeed, our S-ALOHA enables the possibility to double the data exchanged on the network or increase the overall number of sensors.

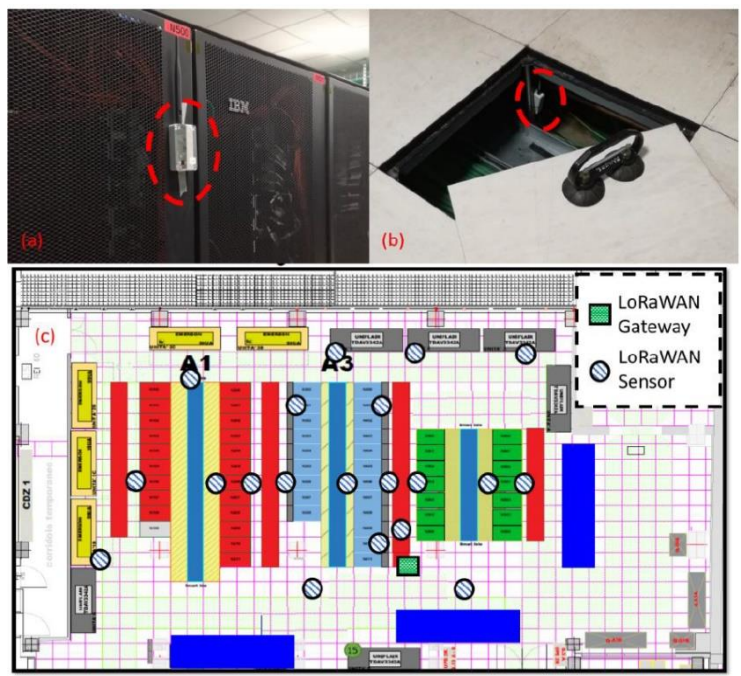

**Figure 11** – Sensor node deployment in CINECA. (a) hallway positioning, (b) the sensor node is under the data center floor, (c) CINECA data center map.

### d. S-ALOHA evaluation

In order to accomplish a performance analysis of the S-ALOHA on LoRaWAN and to make a comparison with the standard LoRaWAN, a detailed log of the status of each device is collected by a database, which is managed by LoRaWAN server. Two packet-send and packet-loss counters were introduced in the state machine indicating the total number of packets that each node sent and those gone lost. The packet-send counter is incremented every time the microcontroller is in the sending state, while to check if the acknowledgment has been received a logical value is set to true a flag present in the receiver function, whenever it assumes logical value true it means that it is received a packet in response, otherwise the S-ALOHA state machine resumes the transmission by increasing the packet-loss counter. Moreover, the gateway detects the channel access time of each end-node, measuring the effective RTC synchronization of the WSN. The payload of each message sent by the end nodes contains the log information, allowing through a debug window to read in real-time the statistics of each node. In order to make the analysis possible, however, it was necessary to memorize this data on a support that can be analyzed remotely with numerical analysis programs such as Matlab or Python.

Tests have been carried out to assess whether the introduction of the S-ALOHA protocol in the LoRa stack leads to an improvement in the overall performance of the network. The LoRaWAN protocol is similar both in terms of channel access and as a transmission algorithm to the P-ALOHA, but it is necessary to consider the fact that the statistical analysis of the P-ALOHA does not take into account the channel occupation for acknowledgment messages. The transmission of the ACK by the gateway in the same channel can, in fact, collide with the other frames that have in the meantime been transmitted by the other devices.

We carried out three tests for the S-ALOHA protocol in three different network traffic conditions with the goal to evaluate the performances gap between the theory, statistical models, and a real deployment. The main variables of these tests are the number of devices as well as the transmission frequency of the messages. Indeed, other parameters, such as payload size and SF are kept constant. In all tests, confirmed messages are used and the same firmware S-ALOHA with a back-off algorithm is implemented in whole devices. The following tables and plots show the comparison between the performance of both protocols in terms of traffic offered and disposed of by the network. These values are temporally normalized, so the results are independent of the duration of the experiment and the overall use of the channel is expressed in percentage value.

**Table 2 -** Summary of the results obtained from the tests with Slotted LoRaWAN

|  | Test 1 | Test 2 | Test 3 |
|---|---|---|---|
| Test duration [s] | 10273 | 6977 | 3524 |
| Packets sent | 2708 | 3285 | 2894 |
| N of end nodes | 10 | 18 | 24 |
| TX period [s] | 15 | 15 | 13 |
| TX success rate [%] | 69 | 55 | 43 |
| Traffic generated (G) | 0.264 | 0.471 | 0.773 |
| Channel throughput (S) | 0.184 | 0.275 | 0.332 |

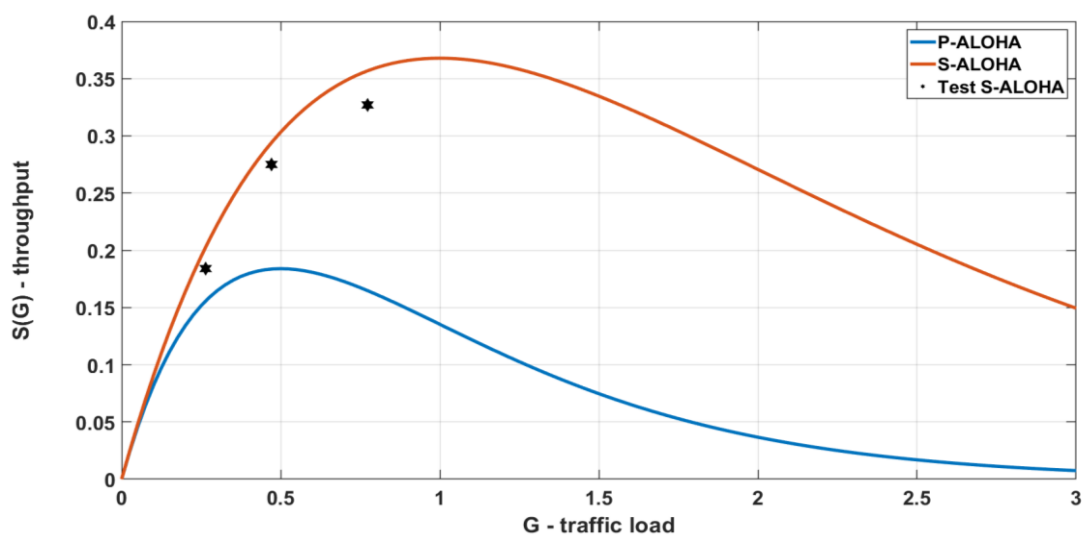

**Figure 12 -** Summary of the results obtained from the tests with Slotted LoRaWAN

For the channel traffic calculation (G) the total number of packets sent in the channel in the time unit (packet/second) was considered; this means that, on average, values greater than G = 1 denote that multiples end nodes transmit simultaneously in the same slots. The overall throughput from the channel is therefore given by the product of G for the average percentage of packages successful disposed of, as described in Equation 7 and 8. The packet time is calculated considering the $T_{LoRaWAN}$.

$$(7) \qquad G = \frac{Packet_{sent}}{T_{test}}$$

$$(8) \qquad S = G * P_{success}$$

Values reported in Table 2 and Figure 12 confirm a correct implementation of the S-ALOHA over a LoRaWAN protocol, indeed the maximum theoretical throughput is doubled, with a trend similar to standard S-ALOHA statistical model.

## 8. Comparison between LoRaWAN and Slotted LoRaWAN

A comparison between the standard LoRaWAN and the S-LoRaWAN is presented in this chapter. The same parameters of the radio configurations and the traffic injected are used for both protocols. In this test the channel traffic and the corresponding throughput are calculated considering only the payload time (T) as traffic generated, the following conditions applied in Equation 6. Moreover, with this approach the tolerance interval ($T_b$) inserted into the slot time is considered into the protocol overhead. The channel throughput in Table 3 describes the effective payload transmitted on the channel; this allows a more accurate comparison without considering the protocol overhead.

**Table 3 -** Comparison between LoRaWAN and Slotted LoRaWAN

|  | Slotted LoRaWAN | LoRaWAN |
|---|---|---|
| Test duration [s] | 3524 | 9596 |
| Packets sent | 2894 | 6817 |
| N of end nodes | 24 | 24 |
| TX period [s] | 13 | 13 |
| TX success rate [%] | 33 | 7 |
| Traffic generated (G) | 0.455 | 0.393 |
| Channel throughput (S) | 0.150 | 0.026 |

Referring to the traffic analysis, Table 3 shows the comparison between the S-LoRaWAN performed with the S-ALOHA MAC (Figure 13 point 1) and the LoRaWAN standard protocol (Figure 13 point 2). Interestingly, the generated traffic resulting from the two protocols is slightly different, this is partly due to the statistical evolution of transmissions; in fact, in the case of the LoRaWAN standard protocol, the overlapping of the frames is much more probable, with consequent recourse to the backoff algorithm and an average reduction in the number of packets transmitted. In the proposed results, the measured throughput of S-LoRaWAN and standard LoRaWAN is respectively 15% and 2.6%. Finally, the measured improvement is 5.8x. This result is higher than the theoretical 2x between S-ALOHA and P-ALOHA; indeed, it has to be considered that the maximum throughput is not placed at the same G (Figure 13), and with the parameters used in this test, the P-ALOHA channel is heavily crowded with a TX success rate of 7%.

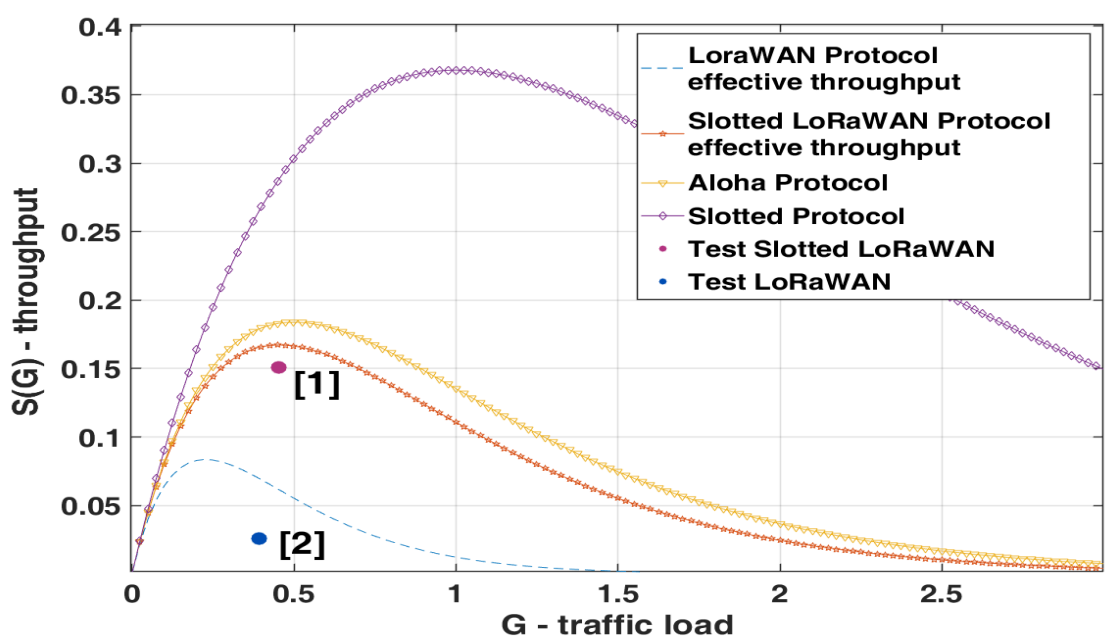

**Figure 13 -** Comparison between LoRaWAN and Slotted LoRaWAN

## 9. Synchronization analysis in Test 3

An in-depth analysis of the synchronization between end nodes in the network is presented here. Test 3 of chapter 7 is used as the reference, indeed the high generated traffic and the high packet collision rate on the channel tresses the synchronization procedure. The error is calculated between $t_0$ (Figure 1 (a)) and the packet time of arrival measured by the gateway. With the $T_b$ proposed in Test 3 the overall number of failed slots for all 24 nodes, is five, with a success rate of 99.6%. In this case, an overlap happens between adjacent slots because the tolerance interval is exceeded. As presented in Figure 14 (a), in point 1 and 2 two consecutive slot overlaps are shown, and both are generated from the same end node. This issue comes from end node reboot; indeed, when a firmware error is detected the MCU is restarted and the internal RTC is not configured. Therefore, the overall synchronization failures are three.

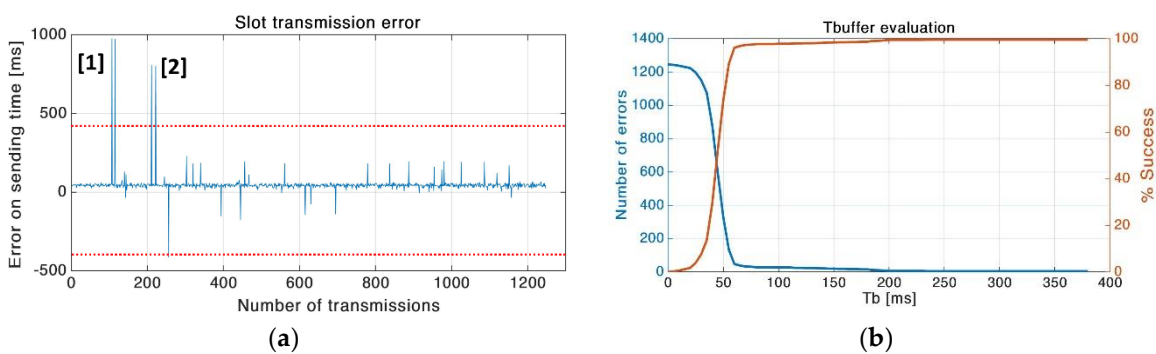

(a)                                                                          (b)

**Figure 14 -** (a) Synchronization error in ms respect to the beginning of each slot. The maximum error allowed is 400 ms, which is exceeded 5 times. Two overflows are generated from end node reboot due to firmware issues; (b) Number of slots overlapped as a function of tolerance interval length.

The previous tests are executed with a tolerance interval ($T_b$) calculated to fulfill the RTC drift in the worst case, we evaluated the S-LoRaWAN throughput in this condition to compare both protocols in a typical low power application. However, in Test 3 the $T_b$ could be decreased ensuring a slot success rate over 90%; in Figure 14 (b) and Table 4 is shown the relationship between $T_b$ and the number of errors. For example, with a $T_b$ of 100ms the error rate is practically the same as Test 3, and with 50ms the error rate is still acceptable. The tolerance interval length affects the channel throughput; indeed, reducing $T_b$ decreases the protocol overhead, and allows more slots in a unit time.

**Table 4 -** Number of slots failed respect to tolerance interval shrinkage

| $T_b$ [ms] | 50 | 100 | 150 | 200 | 250 | 300 |
|---|---|---|---|---|---|---|
| N slot errors | 328 | 27 | 20 | 6 | 5 | 5 |
| Success rate [%] | 74 | 97 | 98 | 99 | 99 | 99 |

## 10. Conclusions

We have presented an analysis of the LoRaWAN protocol focused on channel throughput and issues associated with its scalability. Moreover, we have implemented a reliable synchronization method, used to develop an S-LoRaWAN protocol suitable for low-cost and low-power IoT devices. This approach gives a theoretical 2x network throughput improvement, but in high traffic set-up, the measured improvement is up to 5.8x with a demonstrated reduction of packet collisions of 26% in a real-life deployment with 24 nodes operating for hours. The overall overhead of the proposed S-LoRaWAN is only 8 bytes in the downlink packet, with a negligible consequence in term of power consumption, which is a fundamental factor for battery operated devices. Finally, our S-LoRaWAN does not require any change to the LoRaWAN software stack and can be deployed on top of unmodified LoRaWAN firmware.

**Supplementary Materials:** The LoRa data rate calculator is available online at http://www.rfwireless-world.com/calculators/LoRa-Data-Rate-Calculator.html; the LoRaWAN stack is available online at https://www.st.com/en/wireless-connectivity/lorawan-technology.html.

**Author Contributions:** D.B. proposed the idea of a Slotted LoRaWAN. T.P. and A.M. performed the experiments and the analysis of the results. The article was written by T.P. and D.B., with the assistance of L.B. on the revision of the manuscript.

**Funding:** This research received no external funding.

**Acknowledgments:** The end nodes used to evaluate the S-LoRaWAN in this paper were designed by Emanuele Bedeschi.

**Conflicts of Interest:** The authors declare no conflict of interest.

## Appendix A

The duration of uplink and downlink transmissions depends on parameters of LoRa modulation such as SF (Spreading Factor), BW (Bandwidth), CR (Coding Rate, values from 1 to 4) and can be expressed as the sum of the time needed to transmit the preamble and the physical message (Equation 9).

$$(9) \qquad T_{tx} = T_{preamble} + T_{PHY\ Message}$$

Equations 10 and 11 represent how these two terms have been calculated, where $N_{preamble}$ is the number of symbols used by the radio transceiver as the physical preamble of the message and $N_{PHY}$ indicates the number of symbols transmitted in the physical message and can be determined as shown in Equation 5. $T_{sym}$ (Equation 12) is the duration (in seconds) of a symbol and depends on SF and BW.

$$(10) \qquad T_{preamble} = T_{sym} \cdot \left( N_{preamble} + 4.25 \right)$$

$$(11) \qquad T_{PHY\ Message} = T_{sym} \cdot N_{PHY}$$

$$(12) \qquad T_{sym} = \frac{2^{SF}}{BW}$$

$$(13) \qquad N_{PHY} = 8 + \max \left[ ceil \left[ \frac{28 + 8 \cdot PL + 16 \cdot CRC - 4 \cdot SF}{4 \cdot (SF - 2 \cdot DE)} \right] \cdot (CR + 4), 0 \right]$$

In Equation 13 PL (Payload Length) denotes the number of bytes in the physical payload, CRC indicates the presence (value 1) or not (value 0) of the CRC field in the physical message and DE indicates if the mechanism to prevent issues about the clock drift of the crystal reference oscillator is used (value 1 for SF12 and SF11, 0 for others). The transmission data rate can be obtained as shown in Equation 14:

$$(14) \qquad DR = SF \cdot \frac{BW}{2^{SF}} \cdot \frac{4}{CR + 4}$$